\documentclass{ceab}   

\usepackage{epsfig}     
\usepackage{graphicx}   
\usepackage{url}
\usepackage{natbib}     
\usepackage[T1]{fontenc} 
\usepackage{babel}       
\usepackage{hyphenat}
\usepackage{amssymb,amsmath,amsthm,enumitem}
\usepackage{mathtools}
\usepackage{wrapfig}
\newcommand{\shrink}{\vspace{-0.3cm}}
\usepackage{lineno}

\usepackage{subcaption}

\def\runningtitle{Weyrauch et al.: Event-by-event reconstruction of air-shower events with IceCube}
\def\articlenumber{7}
\def\ppages{1--4}

\begin{document}

\title{Event-by-event reconstruction of air-shower events with IceCube using a two component lateral distribution function}

\author{Mark Weyrauch$^{1}$\thanks{mark.weyrauch@kit.edu} \ for the IceCube Collaboration\thanks{https://authorlist.icecube.wisc.edu/icecube}
\vspace{2mm}\\
\it $^1$Karlsruhe Institute of Technology, Institute for Astroparticle Physics, \\ Karlsruhe, Germany
}

\maketitle

\begin{abstract}
The IceCube Neutrino Observatory, located at the geographic South Pole, comprises a surface component, IceTop, and an optical in-ice array. This unique com\-bi\-na\-tion allows for coincident measurements of low-energy ($\sim \rm{GeV}$) and high-energy ($\gtrsim 400\,\rm{GeV}$) muons produced in cosmic-ray air showers. The ratio of the low- and high-energy muon yields can constitute a useful tool not only for composition analyses but also for testing different hadronic interaction models.
However, since IceTop does not feature dedicated muon detectors, the measurement of the low-energy muon component for individual air showers is challenging. In this work, a new approach for a single-event based estimation of the low-energy muon content using a two component lateral distribution function (LDF) is utilized. This method combines an analytic description for the electromagnetic and muon lateral distributions of the total signal, with an aim to reconstruct primary energy and low-energy muon number, respectively. The underlying principle of this method as well as the resulting reconstruction performance will be detailed in this work.
\end{abstract}


\section{Introduction}
\vspace{-.2cm}
When cosmic rays enter the Earth's atmosphere, they can produce large particle cascades, so called extensive air-showers (EASs), which can be observed via ground-based detector arrays. Muon production in EASs con\-sti\-tutes a significant uncertainty in EASs measurements. To wit, a dis\-crep\-an\-cy between the number of muons in measured data and simulations was reported by some experiments (\cite{Albrecht:2021cxw,Soldin:2021wyv,ArteagaVelazquez:2023fda}). For the derivation of tighter constraints on hadronic interaction models, the IceCube Neutrino Observatory, with its combination of a surface array, IceTop, and deep in-ice detector, offers the unique possibility to correlate low-energy ($\sim$GeV) muons and high-energy ($\gtrsim 400\,\rm{GeV}$) muons. While the GeV muon density has been measured with IceTop in the $2.5-120\,\mathrm{PeV}$ energy range by means of a statistical analysis (\cite{IceCubeCollaboration:2022tla}), for correlation studies an event-by-event estimation is required. 
The approach presented in this work is based on a two-component lateral distribution function (LDF) fit to the signal distribution of EASs observed with IceTop and aims to provide a tool for the reconstruction of the primary energy and low-energy muon number on a single event basis.

\section{IceTop}
\vspace{-.2cm}
IceTop is the square kilometer surface detector of IceCube and comprises 81 detector stations, each equipped with two ice-Cherenkov tanks separated by $\approx10\,\rm{m}$ (\cite{IceCube:2012nn}). The stations are deployed in a triangular grid with a 125\,m spacing, with the exception of a denser infill area focused on low energies ($<50\,$m spacing). Charged particles traversing the ice volume inside a tank create Cherenkov light which is collected by the two digital optical modules (DOMs) in each tank. If the measured signal exceeds the discriminator threshold, a tank has a local trigger (soft local coincidence (SLC)). If both tanks in one station exhibit a local trigger within a $1\,\mu\rm{s}$ time window, the station is classified as a hard local coincidence (HLC). 
The signals measured in IceTop tanks are calibrated to units of vertical equivalent muons (VEM), describing the average charge deposit produced by a muon traversing vertically. In the standard EAS reconstruction (\cite{IceCube:2012nn}), only HLC stations are included to estimate the primary energy and the shower geometry. However, since SLC hits are typically produced by muons, the following analysis includes both HLCs and SLCs in order to additionally reconstruct the muon content with IceTop.

\section{Implementation}
\begin{wrapfigure}{RTH!}{0.58\textwidth}
\vspace{-0.1cm}
 \centering
  \includegraphics[scale=0.5,angle=0]{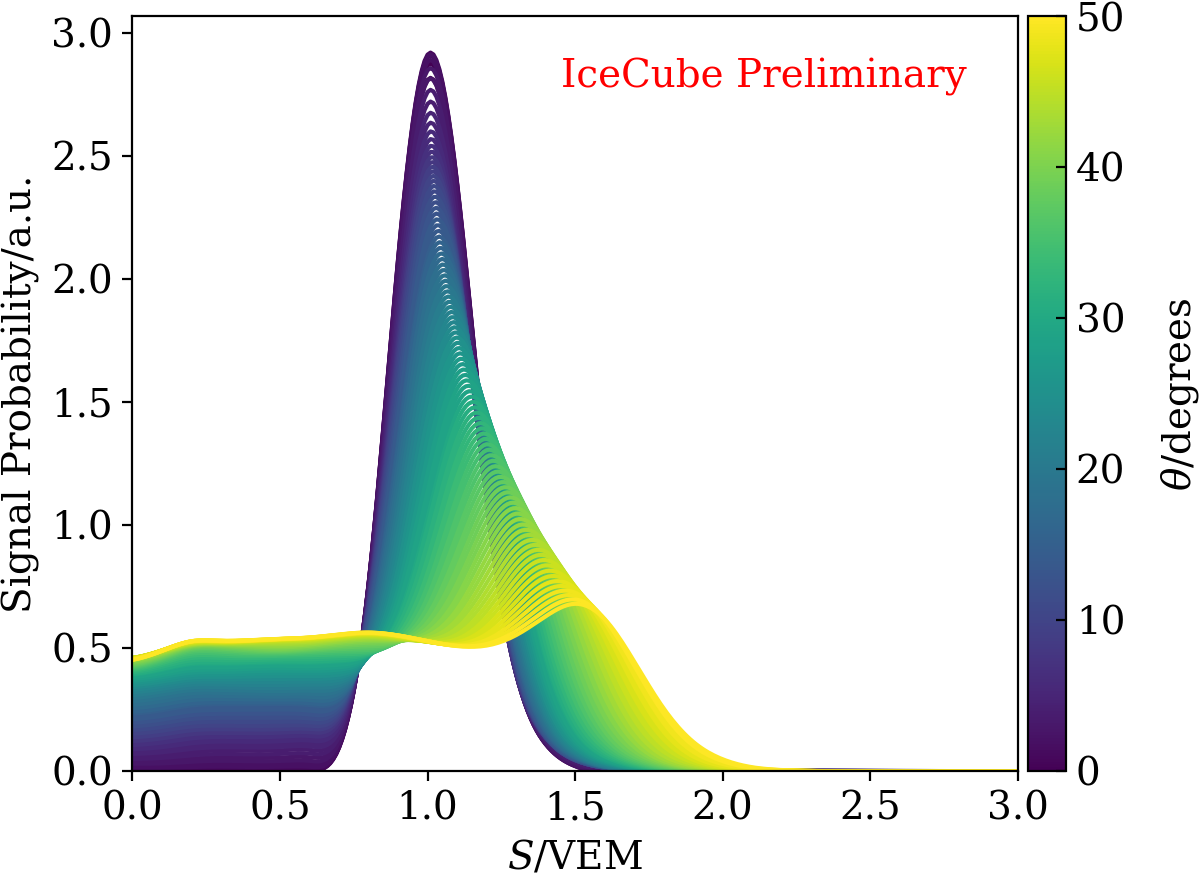}%
 \shrink
 \caption{Spline fits to IceTop tank response simulations for single muon injections as a function of inclination.}
 \label{fig:n1_muon_splines}
 \shrink
\end{wrapfigure}

\begin{figure*}[t!]
    \centering
    \begin{minipage}[c]{0.46\textwidth}
    \centering
    \includegraphics[width=\textwidth]{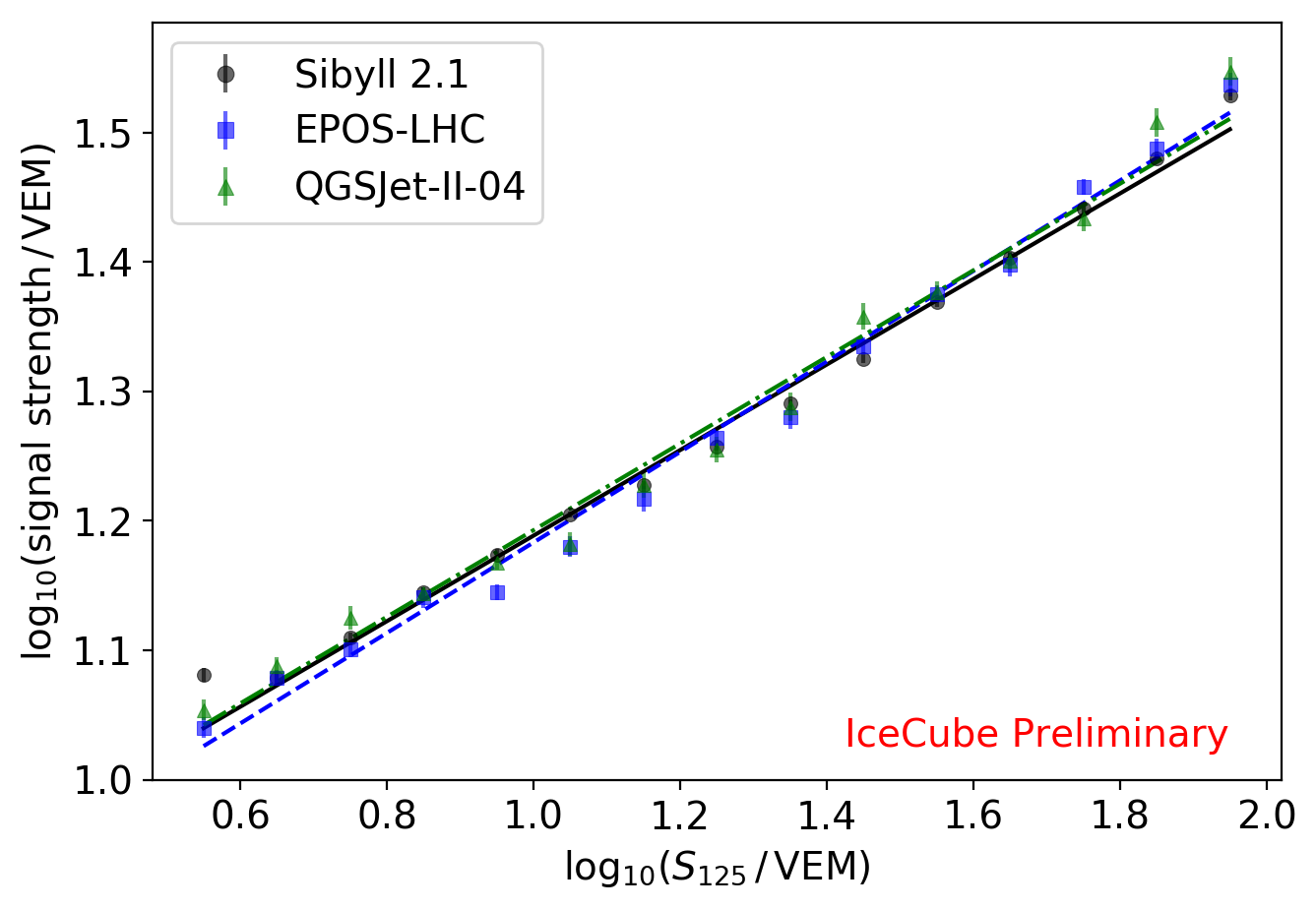}
    \vspace{-0.8cm}
    \caption{Parametrization of the 95\% phase space coverage of the muon contribution as a function of $\log_{10}S_{125}$.}
    \label{fig:95perc_phase_space_fit}
    \end{minipage}
    \hfill
    \begin{minipage}[c]{0.5\textwidth}
    \centering
    \includegraphics[width=\textwidth]{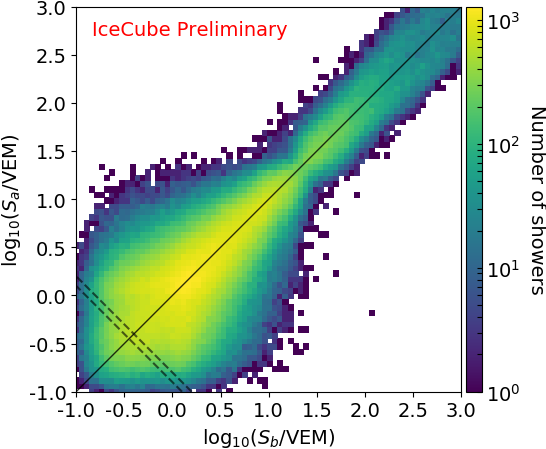}
    \vspace{-0.8cm}
    \caption{Distribution of signal de\-posits in neighboring HLC tanks using 1$\%$ of the 2012 data for $\theta_{\rm{reco}}<26^{\circ}$.}
    \label{fig:neighboring_signals_burnsample}
    \end{minipage}
\shrink
\end{figure*}

\begin{figure*}[t!]
\vspace{-0.5cm}
\shrink
    \centering
    \begin{subfigure}[t!]{0.482\textwidth}
    \centering
    \includegraphics[width=\textwidth]{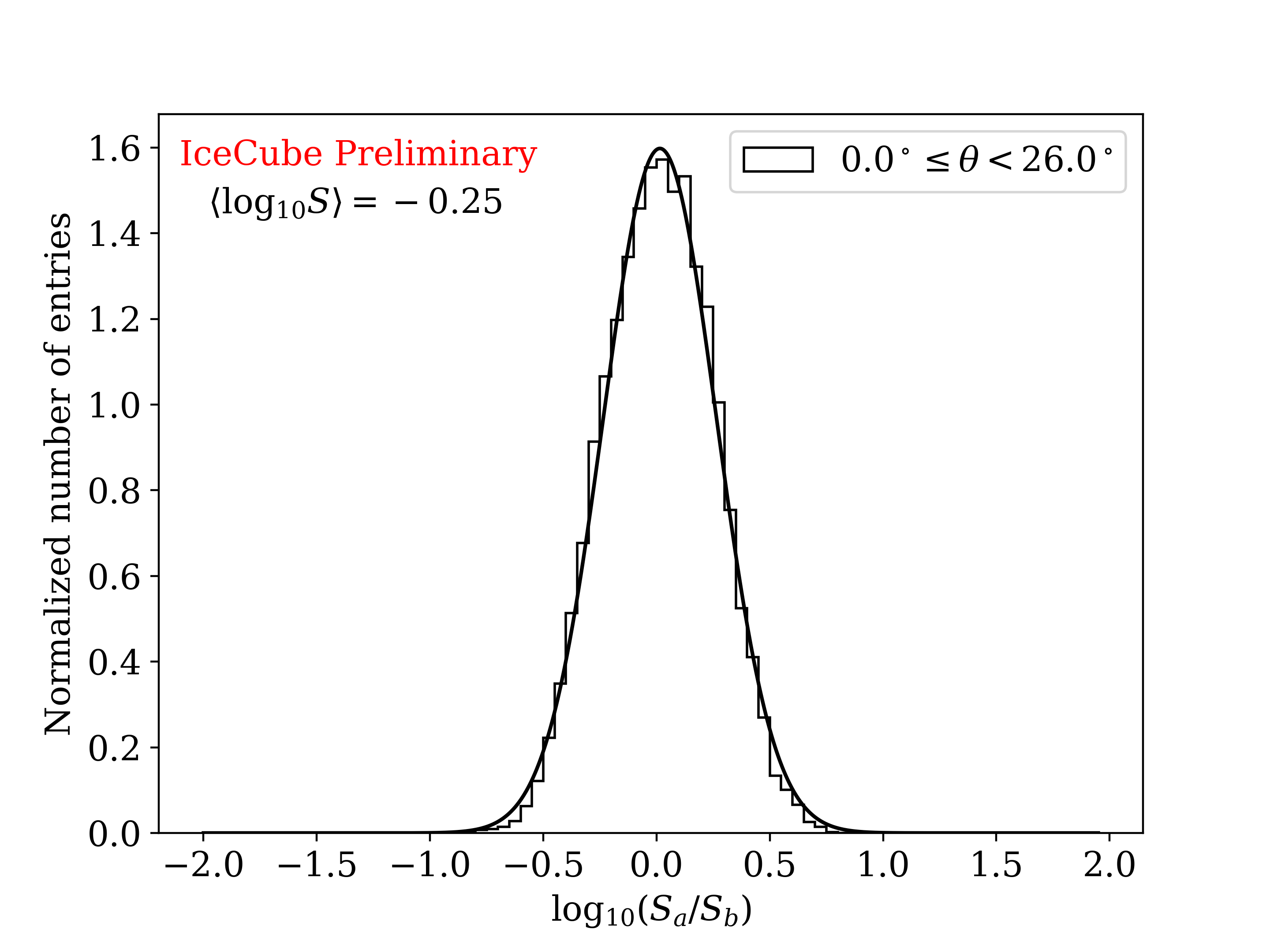}
    \end{subfigure}
    \hfill
    \begin{subfigure}[t!]{0.482\textwidth}
    \centering
    \includegraphics[width=\textwidth]{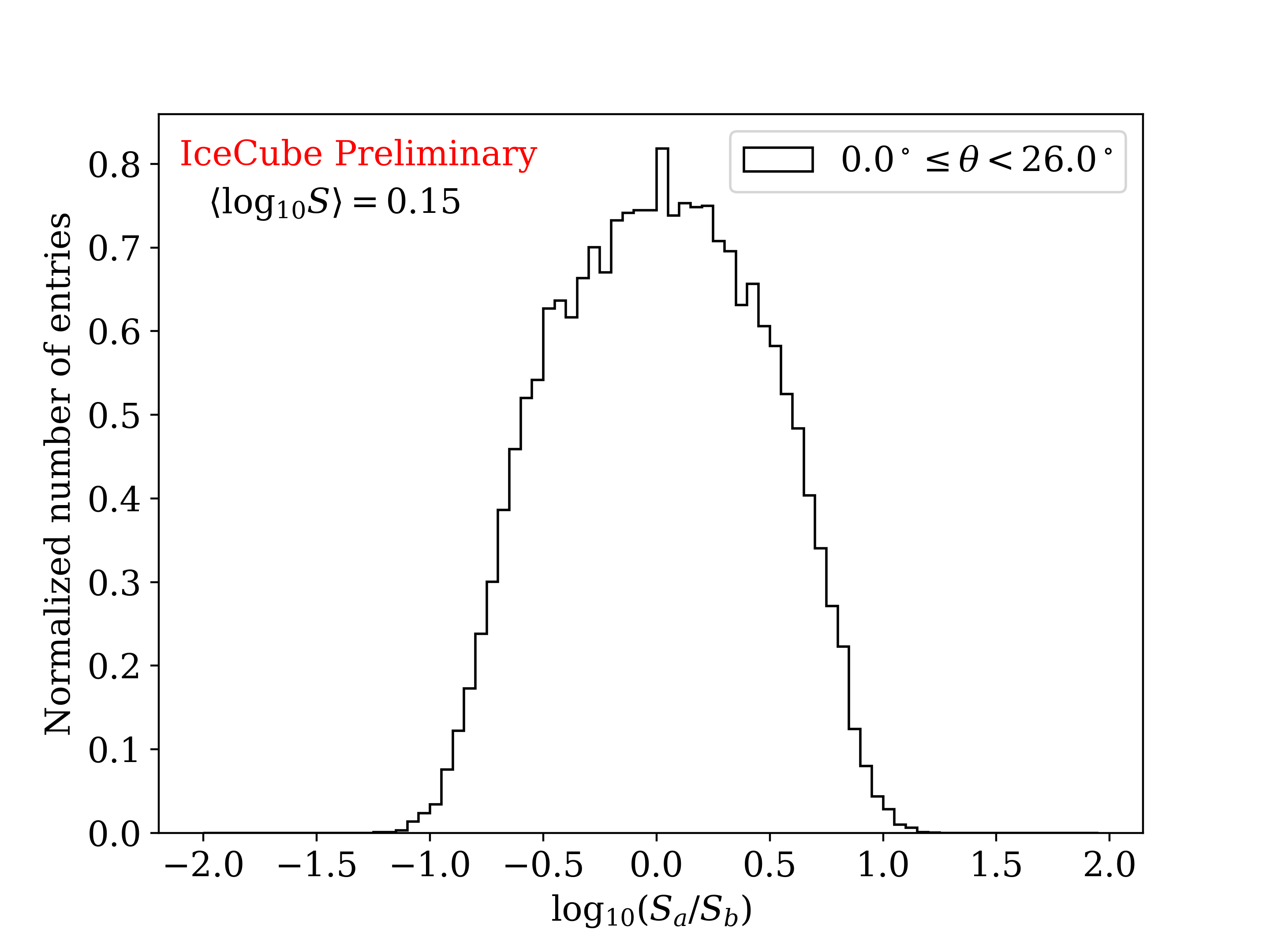}
    \end{subfigure}
    \shrink
    \caption{Example bins of the signal difference in neightboring HLC tanks drawn from Fig. \ref{fig:neighboring_signals_burnsample} for an average signal of $\langle \log_{10}(S/\rm{VEM})\rangle=-0.25$ (left) and $\langle \log_{10}(S/\rm{VEM})\rangle=0.15$ (right).}
    \label{fig:signal_diff_bins}
\shrink \vspace{-0.2cm}
\end{figure*}

\begin{wrapfigure}{R}{0.6\textwidth}
\shrink
\vspace{-0.3cm}
 \centering
  \includegraphics[scale=0.5,angle=0]{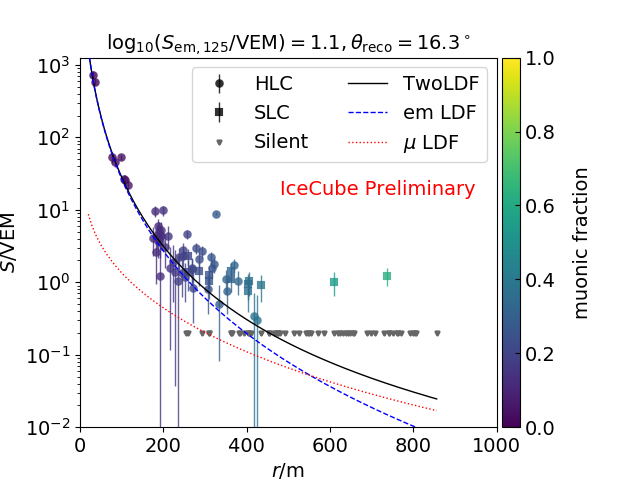}%
 \shrink
 \caption{Two-component LDF fit for a simulated event air-shower event ($\log_{10}(E_{\rm{true}}/\,\rm{eV})=16.59$) in which silent detectors are drawn at a fixed value for visualization purposes. The muon fraction from the fit is indicated by the color scale.}
 \label{fig:single_event_fit}
 \vspace{-0.4cm}
\end{wrapfigure}

\vspace{-.2cm}
\setlength{\belowdisplayskip}{5pt} \setlength{\belowdisplayshortskip}{5pt}
\setlength{\abovedisplayskip}{5pt} \setlength{\abovedisplayshortskip}{5pt}
The lateral distribution functions used for the application within the two-component LDF are the `Double Logarithmic Parabola' (DLP) 
\begin{equation}
    S_{\rm{em}} = S_{\rm{em},125} \left( \frac{r}{r_{\rm{em}}} \right) ^{-\beta_{\rm{em}} - \kappa(S_{\rm{em},125}) \log_{10}{ \left( r/r_{\rm{em}} \right) }} , \ r_{\rm{em}} = 125\,\rm{m} \ , \label{eq:DLP}
\end{equation}
for the electromagnetic (EM) contribution ($e^\pm$,$\gamma$) and a modified NKG function based on the Greisen LDF~(\cite{Greisen:1960wc})
\begin{equation}
    S_{\mu} = S_{\mu,600} \left(  \frac{r}{r_{\mu}} \right)^{-\beta_\mu} \left(  \frac{r+320\,\rm{m}}{r_{\mu}+320\,\rm{m}} \right)^{-\gamma} , \ r_\mu = 600\,\rm{m} \ , \label{eq:Greisen}
\end{equation}
for the muon contribution. Each function has two slope parameters de\-scrib\-ing the shape of an observed charge distribution. For the DLP function, the parameters $\beta_{\rm{em}}$ and $\kappa$ describe the slope and curvature, respectively. While the former is a free parameter within the reconstruction, the curvature is parametrized as a function of $S_{\rm{em},125}$ based on average EM LDFs. In case of the muon LDF, the LDF shape is determined by $\beta_\mu$ and $\gamma$. In order to derive the primary energy and the low-energy muon number, both functions feature a respective proxy parameter, $S_{\rm{em},125}$ and $S_{\mu,600}$. Both proxy parameters correspond to the expected signal strength given by the LDF fit at a reference distance relative to the shower axis. For the energy reconstruction, a reference distance of $125\,\rm{m}$ is found to provide a com\-bi\-na\-tion of a small primary mass de\-pen\-dence and a good reconstruction resolution (\cite{Andeen:2011}). The proxy parameter for the muon number is taken at a distance of $600\,\rm{m}$ in order to minimize the influence of the EM contribution while limiting the effect of the large fluctuations present far from the shower axis. 
While the air-shower reconstruction via the DLP function is well known from standard IceTop analyses (\cite{IceCube:2012nn}), the integration of the modified NKG function (\ref{eq:Greisen}) describing the muon contribution is novel. 
The likelihood function for the EM contribution is based on a log-normal distribution as parametrized in~\cite{IceCube:2012nn}.
\\ \noindent
For the derivation of the likelihood function for the muon contribution, dedicated IceTop tank re\-sponse si\-mu\-la\-tions are uti\-lized. For this purpose, the IceTop tank response for injected ($E_\mu=1\,\rm{GeV}$) muons with given multiplicity and in\-cli\-na\-tion is simulated with Geant4 (\cite{GEANT4:2002zbu}). The resulting signal response PDF for single muons as a function of their inclination (calibrated to units of VEM) is shown in Fig. \ref{fig:n1_muon_splines}. It can be noted, that the signal response is a function of the detector geometry. As the inclination in\-creas\-es, the peak position shifts as a function of $1/\cos{\theta}$, while the con\-tri\-bu\-tion of muons only tra\-ver\-sing the edge of the detector (`corner clipping muons') in\-creas\-es simultaneously. The signal spread therefore in\-creas\-es with muon in\-cli\-na\-tion. The resulting muon signal PDFs are then saved as spline fits in order to retrieve the probability val\-ues $p_{\mu,\rm{sig}}(S|\theta, n)$ 

\noindent for a given muon in\-cli\-na\-tion, $\theta$, mea\-sured signal strength, $S$, and muon multiplicity, $n$, during the re\-con\-struc\-tion. Since the muon in\-cli\-na\-tion can not be mea\-sured with the IceTop tanks, the muons are approximated to follow the primary zenith angle. The effect of a cor\-rec\-tion for the muon zenith angles is currently under investigation. The muon sig\-nal PDFs are then weighted with a Poisson factor based on the average expected number of muons
\begin{equation}
    p_\mu(S|\theta, \langle N_\mu \rangle) = \sum_{n}^{} \frac{\langle N_\mu \rangle^n}{n!} e^{-\langle N_\mu \rangle}  p_{\mu,\rm{sig}}(S|\theta,n) \ , 
    \label{eq:muPDF}
\end{equation}
as determined by the product of the expected muon signal $\langle S_\mu \rangle$ (given by the muon LDF) and the effective tank area $A_{\rm{eff}}$. For large muon multiplicities ($n>15$), the muon signal PDF can be approximated by a Gaussian. 
\\ \\ \noindent
The combined total probability, taking into account additionally the attenuation of the EM signal caused by the snow accumulation on top of the IceTop tanks, is calculated as
\begin{gather}
\begin{aligned}
p_{\text{SLC}}\left( S|\theta , \langle S_{\rm{em}} \rangle, \langle S_{\mu} \rangle \right) \hspace{-2pt} &= \hspace{-2pt} \int_{0}^{S} \hspace{-8pt} p_{\rm{em}}(S_{\rm{em}}'|\theta,\langle S_{\rm{em}}\rangle c_{\rm{snow}}) p_{\mu}(S-S_{\rm{em}}'|\theta, \langle N_\mu \rangle) dS_{\rm{em}}' \, , \\
p_{\text{HLC}}\left( S|\theta , \langle S_{\rm{em}} \rangle, \langle S_{\mu} \rangle \right) \hspace{-2pt} &= \hspace{-2pt}
\begin{dcases}
     p_{\rm{em}}(S-\langle S_{\mu} \rangle|\theta,\langle S_{\rm{em}}\rangle c_{\rm{snow}}), & S > f_{\mu}(S_{125}) \, , \\
     e^{-\langle N_\mu \rangle} p_{\rm{em}}(S|\theta,\langle S_{\rm{em}}\rangle c_{\rm{snow}}) , & \log_{10} S < -0.2 \, , \\[-4pt]
     \int_{0}^{S} p_{\rm{em}}(S_{\rm{em}}'|\theta,\langle S_{\rm{em}}\rangle c_{\rm{snow}}) \\[-8pt] \quad \quad p_{\mu}(S-S_{\rm{em}}'|\theta, \langle N_\mu \rangle) dS_{\rm{em}}' , & \text{else} \, ,
  \end{dcases}
\end{aligned} \label{eq:p_signal}
\raisetag{14pt}
\end{gather}
for SLC and HLC hits, respectively. This separate treatment is motivated by the dif\-fer\-ence in the relative con\-tri\-bu\-tion of EM par\-ti\-cles and muons de\-pend\-ing on the observed signal strength. Since for single tank triggers a
\begin{wrapfigure}{RHT!}{0.64\textwidth}
\vspace{-0.1cm}
 \centering
  \includegraphics[scale=0.35,angle=0]{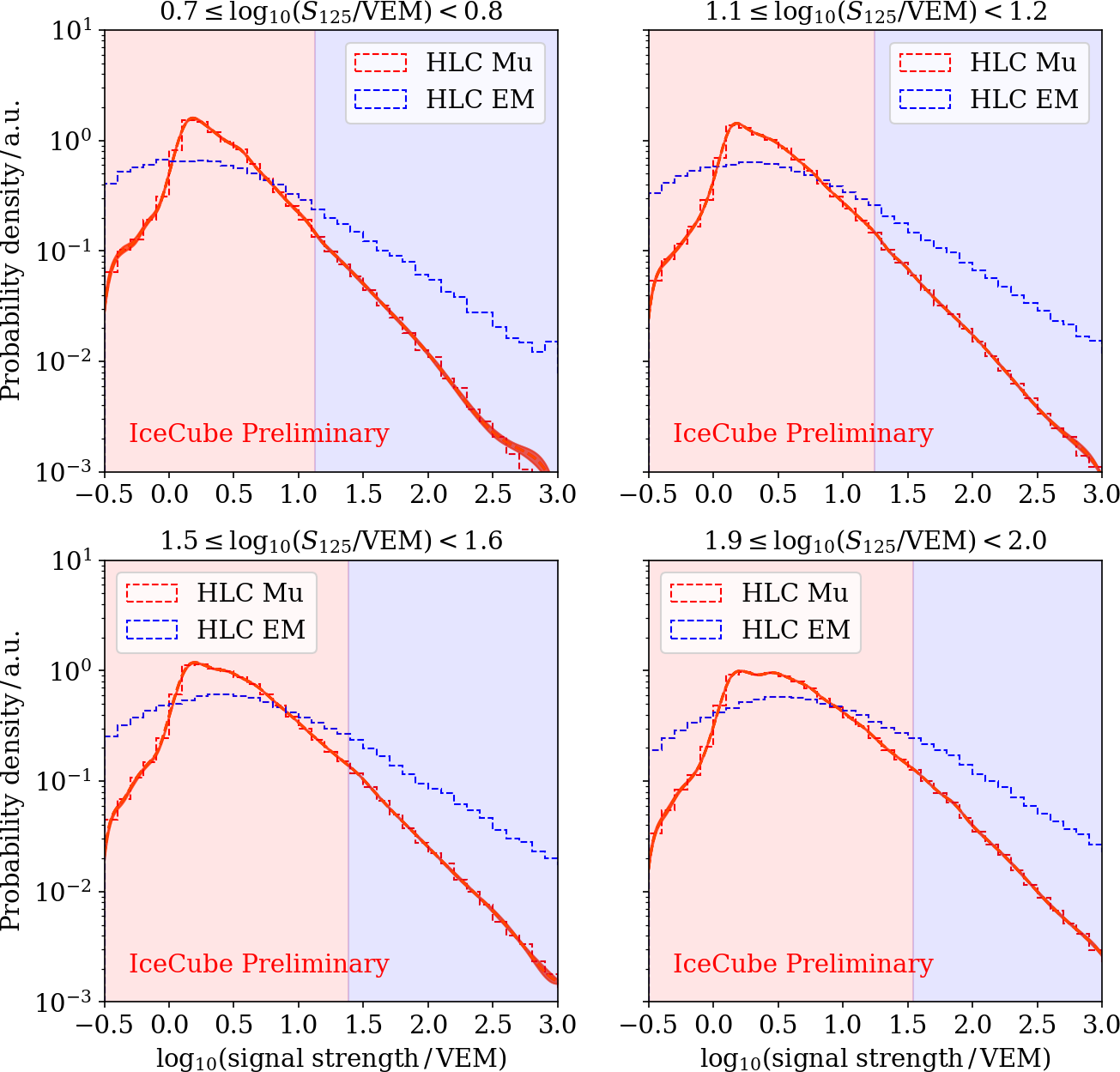}%
 \shrink
 \caption{PDFs for muons \& electromagnetic particles from air-shower simulations produced with Sibyll 2.1 as a function of the cumulative charge measured in full station triggers (HLCs).}
 \centering
 \shrink \vspace{-0.1cm}
 \label{fig:hlc_muon_pdfs}
\end{wrapfigure}

\noindent
thresh\-old of $0.7\,\rm{VEM}$ is applied to only include hits that likely have significant muon con\-tri\-bu\-tion (\cite{ICRC23WeyrauchSoldin}), the EM and muon PDFs are always fully con\-volved for SLCs. In the case of HLC hits, however, a distinction of the regions in which the footprint is dominated by EM par\-ti\-cles is im\-por\-tant for the per\-for\-mance of the re\-con\-struc\-tion. Firstly, in the region close to the shower axis, muons only contribute a small frac\-tion to the measured signal strength. In this region, only the EM contribution is taken into account in the like\-li\-hood. In order to find a proper clas\-si\-fi\-ca\-tion for this regime, the phase space of the average muon contribution as a function of the total charge measured in HLC hits (Fig. \ref{fig:hlc_muon_pdfs}) is studied using Monte-Carlo simulations. Since the relative contribution of muons to a measured signal depends on the shower size, the distributions are produced in bins of the standard energy proxy $\log_{10}S_{125}$ (\cite{IceCube:2012nn}). 
The normalized and H4a (\cite{Gaisser:2011klf}) weighted muon distributions in each bin are described with a spline fit and integrated up to a threshold of 0.95 in order to account for 95\% of the phase space, indicated as red shaded region. This value results in a good overall performance in energy and muon number reconstruction. 
The corresponding signal strength values as a function of  $\log_{10}S_{125}$ are shown in Fig. \ref{fig:95perc_phase_space_fit} and show only small differences among different hadronic interaction models. The resulting parametrization, $f_{\mu}(S_{125})$, based on Sybill 2.1 is used in the reconstruction procedure.
Secondly, a small signal regime for HLCs in which, again, the average signal is dominated by EM particles, can be distinguished. For this purpose, signal deposits in neighboring tanks of HLC stations can be utilized (Fig. \ref{fig:neighboring_signals_burnsample}). To prohibit any additional model de\-pen\-dence in the classification of the small signal regime, the distribution is drawn from experimental data 
\noindent (1\% of the year 2012) for $\log_{10}S_{125}\geq0.5$. For a better visualization, two example bins of the signal difference between neighboring tanks are shown in Fig. \ref{fig:signal_diff_bins}. For small signal deposits ($\log_{10} S \lesssim -0.2$), the difference distribution can be well described by a log-normal distribution. The underlying signal distribution for HLC hits in this signal regime is therefore log-normal and, thus, on average dominated by EM particles. It should be noted, that the given threshold value is merely a rough estimate and its exact value and implementation can be adjusted for optimization purposes. The impact of the exact choice of the threshold value on the reconstruction performance is currently under investigation. The log-normal EM PDF in the small signal regime is additionally weighted with a factor of $e^{-\langle N_\mu \rangle}$, accounting for the probability to not observe a muon for a particular hit, given the average expected muon number, $\langle N_\mu \rangle$, as determined by the muon LDF. For slightly larger signal deposits roughly above 1\,VEM, the distribution is clearly deviating from a log-normal behavior, indicating, on average, a significant muon contribution. Therefore, in the intermediate region between the small and large HLC signal regimes, a full convolution of the EM and muon contribution is included in the calculation of the likelihoods.
\\ \noindent The total log-likelihood is given as 
\begin{gather}
\begin{aligned}
   \mathrm{llh} &= \log\left( p_{\text{HLC}}\right) + \log\left( p_{\text{SLC}}\right) + \mathrm{llh_{sat}} + \mathrm{llh_{sil}} + \mathrm{llh_{t}} \\
   \rm{llh_{sil}} &= \log\left( 1 - p_{\rm{nohit}}^n \right) , \ n=2/1 \ \text{if silent HLC/SLC} \ . \label{eq:tot_llh}
\end{aligned}
\end{gather}
The saturation and time likelihood ($\mathrm{llh_{sat}}$, $\mathrm{llh_{t}}$) are described in~\cite{IceCube:2012nn}. Non-triggered (`silent') tanks are incorporated in $\mathrm{llh_{sil}}$. The corresponding no-hit probability $p_{\rm{nohit}}$ is derived by integrating up to the threshold value, $s_{\rm{thr}}$, as
\begin{gather}
\begin{aligned}
p_{\rm{nohit}} &=
\begin{dcases}
\int_{S_{\rm{em}}=0}^{s_{\rm{thr}}}  e^{-\langle N_\mu \rangle}p_{\rm{em}}(S_{\rm{em}}'|\theta,\langle S_{\rm{em}}\rangle c_{\rm{snow}}) , \hspace{60pt} \log_{10} S < -0.2 \, ,  \\
\hspace{-2pt} \sum_{n}^{} \hspace{-2pt} \frac{\langle N_\mu \rangle^n}{n!} e^{-\langle N_\mu \rangle} \hspace{-8pt} \int\displaylimits_{S_{\mu}=0}^{s_{\rm{thr}}} \hspace{-3pt} \int\displaylimits_{S_{\rm{em}}=0}^{s_{\rm{thr}}-S_{\mu}} \hspace{-14pt} p_{\mu,\rm{sig}}(S_{\mu}'|\theta,n)  p_{\rm{em}}(S_{\rm{em}}'|\theta,\hspace{-2pt} \langle S_{\rm{em}} \rangle c_{\rm{snow}}) , \text{else} \, , 
\end{dcases} \\
\end{aligned}
\raisetag{15pt}
\end{gather}
in which the small signal regime is included analogously to Eq. \ref{eq:p_signal}. 
\noindent The re\-con\-struc\-tion framework in\-cor\-po\-rat\-ing the negative log-likelihood minimization pro\-ce\-dure is described in \cite{Lesz:2023icrc}. The starting point for the 3-step two-com\-po\-nent LDF re\-con\-struc\-tion is a standard IceTop re\-con\-struc\-tion, which provides high quality seeds for the shower geometry as well as for $S_{\rm{em},125}$~(\cite{ICRC23WeyrauchSoldin}). As a first step, the shower core position as well as $\beta_{\rm{em}}$, $S_{\mu,600}$, $\beta_\mu$ and $\gamma$ are varied in order to adjust the muon LDF. In the following two steps, the shower geometry and the parameters of both LDFs (except for $\kappa$) are varied successively. An example of a two-component LDF fit for a single event as obtained after the last step is shown in Fig. \ref{fig:single_event_fit}, where the black line indicates the sum of both components.

\section{Reconstruction performance}
\vspace{-.2cm}

\begin{figure*}[t!]
    \vspace{-0.5cm}
    \centering
    \begin{subfigure}[h!]{0.495\textwidth}
    \centering
    \includegraphics[width=\textwidth]{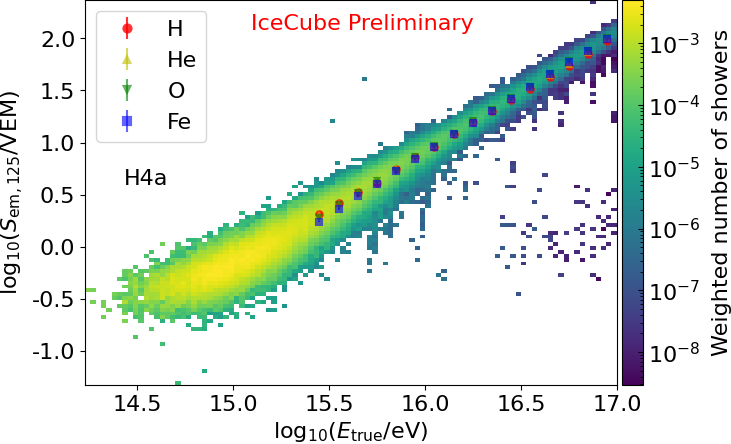}
    \end{subfigure}
    \hfill
    \begin{subfigure}[h!]{0.495\textwidth}
    \centering
    \includegraphics[width=\textwidth]{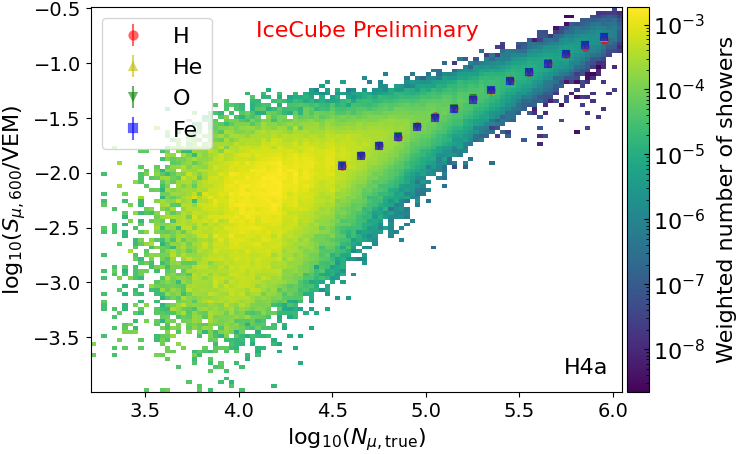}
    \end{subfigure}
    \begin{subfigure}[h!]{0.495\textwidth}
    \centering 
    \includegraphics[width=\textwidth]{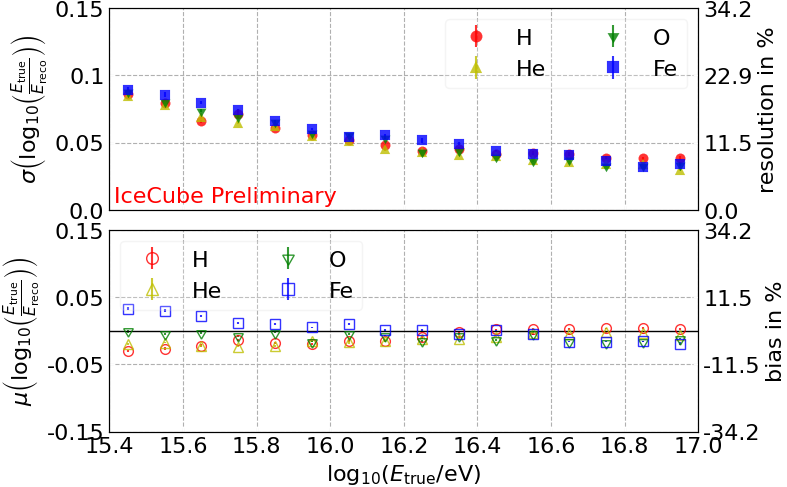}
    \end{subfigure}
    \hfill
    \begin{subfigure}[h!]{0.495\textwidth}
    \centering 
    \includegraphics[width=\textwidth]{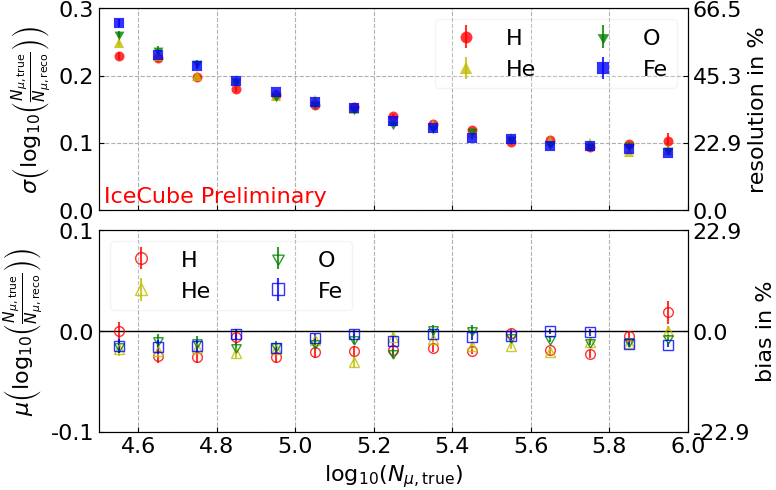}
    \end{subfigure}
    \shrink
    \caption{Upper row: Weighted distribution of the energy proxy, $S_{\rm{em},125}$ (left), and the muon number proxy, $S_{\mu,600}$ (right), for $\theta_{\rm{reco}}<26^\circ$. Mean values for different species are shown approximately in the regime where the IceTop trigger and event selection is fully efficient. Lower row: Corresponding reconstruction resolution and bias in the upper and lower panels, respectively.}
    \label{fig:reconstruction_performance}
\shrink \shrink
\end{figure*}

\begin{wrapfigure}{R!}{0.55\textwidth}
 \centering
  \includegraphics[scale=0.44,angle=0]{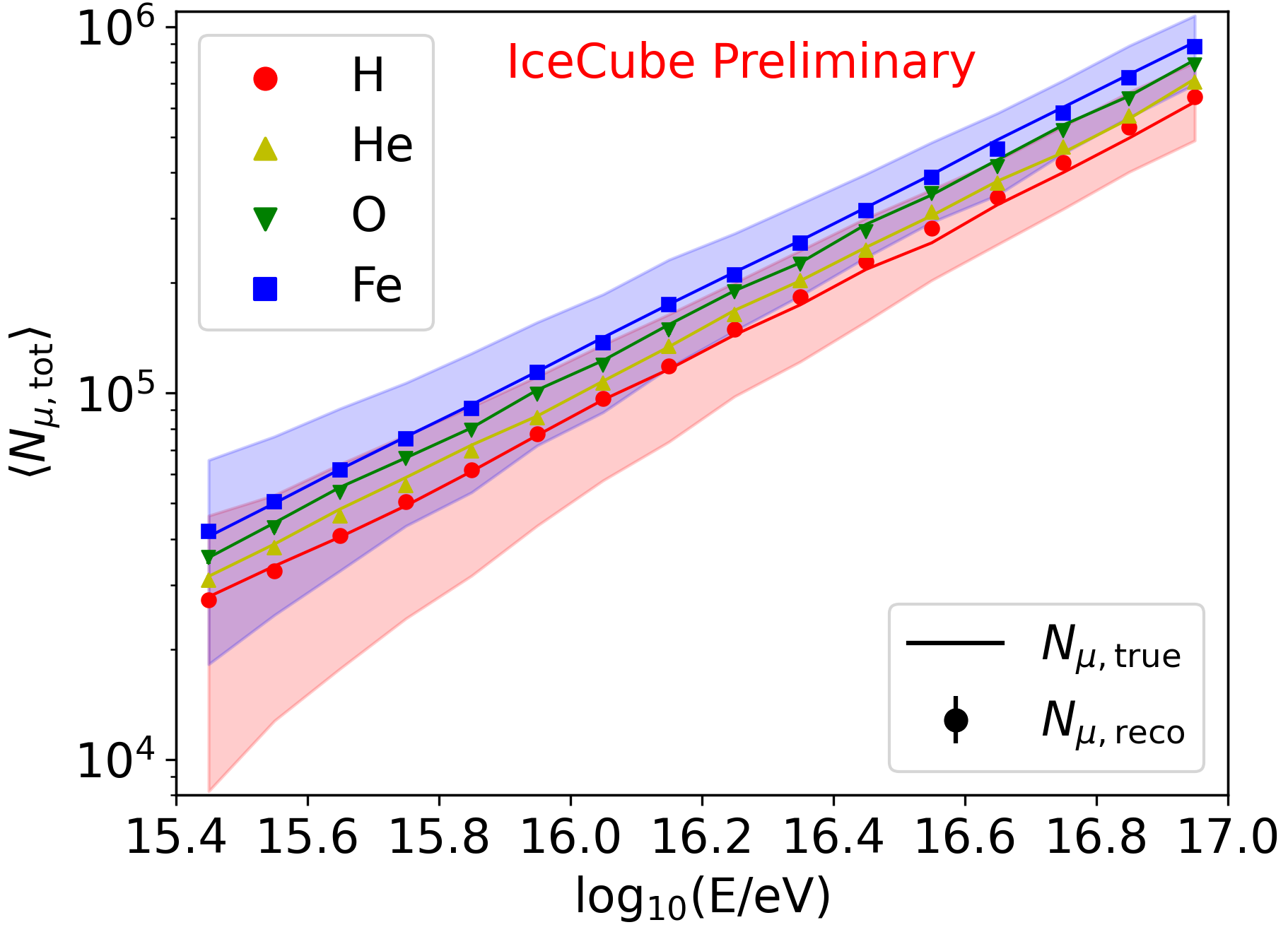}%
 \shrink \vspace{-0.05cm}
 \caption{Average muon number as a function of reconstructed energy. Lines indicate the Monte-Carlo truth.}
 \label{fig:avg_muon_number_vs_energy}
 \vspace{-0.5cm}
\end{wrapfigure}

For the evaluation of the reconstruction performance using the two-com\-po\-nent LDF, the reconstruction is applied to a simulation set for the year 2012, produced with Sibyll 2.1 (\cite{Sibyll2.1:2009}), which contains air-show\-er events for proton, helium, oxygen and iron primaries with energies of $100\,\rm{TeV}-100\,\rm{PeV}$. For the re\-con\-struc\-tion of the pri\-ma\-ry energy and the muon num\-ber, a set of standard quality cuts as discussed in~\cite{Aartsen:2013} is applied. Ad\-di\-tion\-al\-ly, only quasi-vertical showers ($\theta \lesssim 26^\circ$) are included in this analysis. The distribution of the proxy parameters $S_{\rm{em},125}$ and $S_{\mu,600}$ as a function of the corresponding true quantities ($N_{\mu,\rm{true}}=N_{\mu,\rm{true}}(E_{\mu}>210\,\rm{MeV})$) is shown in Fig. \ref{fig:reconstruction_performance} (upper row). The corresponding re\-con\-struc\-tion resolution and bias is shown in the lower row. While the primary energy can be reconstructed with only a small primary mass dependence (few percent level), the re\-con\-struc\-tion of the muon number exhibits no significant dependence on the primary mass. In the range above $\sim 10\,$PeV, the two-component LDF model allows for an energy re\-con\-struc\-tion on the $\sim 12\%$ level and improves to below $\sim 10\%$ toward higher energies. The spread in the muon number is expectedly larger, translating to a re\-con\-struc\-tion resolution of $\sim 20$\% for the highest energies con\-sid\-ered. The average reconstructed muon number as a function of the reconstructed primary energy compared to the Monte-Carlo truth is shown in Fig. \ref{fig:avg_muon_number_vs_energy}, where the reconstructed muon number is shifted by an average correction factor to account for the small re\-con\-struc\-tion offset (see Fig. \ref{fig:reconstruction_performance}). While the error bars of the points indicate the standard error of the mean, the corresponding standard deviations for proton and iron showers are shown as a shaded area.

\vspace{-1cm}
\section{Conclusion}
\vspace{-.2cm}
In this work it was shown, that a combined fit of two model functions for the electromagnetic and muon contribution to the measured charge deposit can be utilized to reconstruct both the primary energy and the muon number on single event basis in cosmic-ray air shower events measured with IceTop. 
For the re\-con\-struc\-tion, the signal expectations at a reference distance of 125\,m and 600\,m as determined by the electro\-magnetic and muon LDF, respectively, can be used as proxy parameters. 
Both primary energy and muon number can be reconstructed with minimal dependence on the primary particle species. While the re\-con\-struc\-tion resolution for the latter reaches $\sim 20\%$ around 100\,PeV, the primary energy can be reconstructed at below 10\% resolution. 
Due to the accumulation of snow on top of the IceTop tanks and consequent attenuation of the electromagnetic signal, the ability to isolate the muon contribution is expected to increase with time. Thus, the application of the two-component LDF to the measured data of

\noindent
different years is of high in\-ter\-est for future analyses.
Gen\-er\-al\-ly, for tests of hadron\-ic in\-ter\-ac\-tion mod\-els and for composition stud\-ies, a single event based muon es\-ti\-ma\-tor con\-sti\-tutes an im\-por\-tant pa\-ra\-me\-ter, in par\-tic\-u\-lar,  in combination with cor\-re\-spond\-ing in\-for\-ma\-tion about the high-energy muon content of the same showers (\cite{verpoest2023multiplicitytevmuonsair}). Ad\-di\-tion\-al\-ly, the slope of the muon LDF con\-sti\-tutes an entirely new parameter for model tests and will therefore be studied in more detail.

\vspace{-.2cm}
\bibliographystyle{apalike}
\bibliography{ECRS24}

\end{document}